\documentclass[12pt]{iopart}

\usepackage{iopams}
\usepackage{graphicx,nicefrac} 
\usepackage{cite}

\begin{document}
\title[Charged fluids encircling compact objects]{Charged fluids encircling compact objects: force representations and conformal geometries}

\author{Ji\v{r}\'{\i} Kov\'{a}\v{r}$^1\!\!$, Yasufumi Kojima$^2$, Petr Slan\'{y}$^1$, Zden\v{e}k Stuchl\'{\i}k$^1$ and Vladim\'{\i}r Karas$^3$}
\address{
$^1$~Research Centre for Theoretical Physics and Astrophysics, Institute of Physics, Silesian University in Opava, Bezru\v{c}ovo n\'{a}m.~13, 746\,01~Opava, Czech~Republic\\
$^2$~Department of Physics, Hiroshima University, 739-8526 Higashi-Hiroshima, Japan\\
$^3$~Astronomical Institute, Czech Academy of Sciences, Bo\v{c}n\'{i} II, 141\,31~Prague, Czech~Republic}
\ead{jiri.kovar@physics.slu.cz, ykojima-phys@hiroshima-u.ac.jp}

\begin{abstract}
Charged fluids rotating around compact objects can form unique equilibrium structures when ambient large-scale electromagnetic fields combine with strong gravity.
Equatorial as well as off-equatorial toroidal structures are among such figures of equilibrium with a direct relevance for astrophysics. To investigate their geometrical shapes and physical properties in the near-horizon regime, where effects of general relativity play a significant role, we commonly employ a scheme based on the energy-momentum conservation written in a standard representation.
Here, we develop its interesting alternatives in terms of two covariant force representations, both based on a hypersurface projection of the energy-momentum conservation. In a proper hypersurface, space-like forces can be defined, following from a decomposition of the fluid four-acceleration.
Each of the representations provides us with an insight into properties of the fluid flow, being well reflected in related conformal hypersurface geometries; we find behaviour of centrifugal forces directly related to geodesics of these conformal hypersurfaces and their embedding diagrams. We also reveal correspondence between the charged fluid flow world-lines from an ordinary spacetime, and world-lines determined by a charged test particles equation of motion in a conformal spacetime.
\end{abstract}


\maketitle

\section{\label{Sec:Intro}Introduction}
Investigation of fluids rotating in large-scale gravitational and electromagnetic fields belongs to problems that have been discussed in physics for decades. In scenarios of astrophysical significance, fluids manifest typically either as an electrically neutral gas, but more-likely as quasi-neutral plasma, otherwise as a neutral or charged microscopic dust, dust grains (pressure-less fluid), and very often as a dispersed medium, such as a dusty-grain gas, dusty plasma, etc. Near compact objects,  astrophysical fluids can impressively whirl within accretion processes, possibly combined with a creation of jets, both characteristic for accretion discs.     
On the other hand, here, we can also find fluids in stable figures of equilibrium, particularly those forming rotating toroidal structures (thick discs) with negligible loss of mass. 

Assuming mean free paths of fluid particles much smaller than a characteristic size of a studied structure, i.e. collisional frequencies between particles much greater than orbital and oscillatory frequencies of the system, we can relax a kinetic theory description for an investigation of the charged fluids, and conveniently use simpler magneto-hydrodynamic equations  \cite{Pun:2008,Rez-Zan:2013}. 
However, even these equations can be mostly solved only by using numerical methods, and simplifying assumptions must be considered. These are, particularly, the well-known assumptions of infinite electric conductivity of the fluid -- the approximation of ideal magneto-hydrodynamics (magnetic field lines frozen in the fluid), and quasi-neutrality of the fluid, reasonable in many astrophysical scenarios involving plasma motion.\footnote{On the other hand, even in more complicated cases when effects of finite conductivity are considered \cite{Koi:2010,Pal-etal:MNRAS-2009,Kud-Kab:1996}, some semi-analytic methods can be successfully applied for studies of particular problems \cite{Lov-etal:APJS-1986,Pra-etal:JAA-1989,Tri-etal:MNRAS-1990,Ban-etal:APJ-1997}.}

However, unlike the limit of ideal magneto-hydrodynamics and plasma consideration, it turns out to be interesting to address a different simplified scenario, which can be handled semi-analytically, or even analytically in special cases. It is the scenario of a fluid with global non-zero charge and pure convective transport of charge across background magnetic field lines. 
Within this approximation, a general relativistic model for the rotating charged fluid can be constructed and tested in various scenarios \cite{Kov-etal:PRD-2011,Kov-etal:PRD-2014,Kov-etal:PRD-2016,Tro-etal:PRD-2018,Schr-etal:PRD-2018,Stu-etal:Uni-2020,Tro-etal:PRD-2020}. By choosing proper background gravitational and electromagnetic fields, i.e. a spacetime geometry metric field and Faraday electromagnetic tensor field, together with a rotational regime of the fluid and its charge distribution, several unique equilibrium structures can be found. Along with the pure equatorial toroidal structures, characteristic mainly for a neutral fluid rotation \cite{Abr-Jar-Sik:AA-1978,Koz-Jar-Abr:AA-1978,Stu-Sla-Hle:AA-2000,Fon-Dai:MNRAS-2002,Rez-Zan-Fon:AA-2003,Sla-Stu:CQG-2005,Stu-Sla-Kov:CQG-2009,Pug-Stu:APJS-2015,Pug-Stu:APJS-2016,Pug-Stu:APJS-2017}, also the off-equatorial ones, the so-called `levitating tori' \cite{Kov-etal:PRD-2016}, or the structures hovering above the central object, referred to as `polar clouds' \cite{Kov-etal:PRD-2014,Tro-etal:PRD-2018,Schr-etal:PRD-2018}, can exist. 

Particularly, the reason for a consideration of this charged fluid approximation was an idea to construct the levitating tori as a generalization of the so-called `halo orbits' -- the orbits of charged test particles rotating in given gravitational and electromagnetic fields with fixed off-equatorial latitudes \cite{Kov-Stu-Kar:CQG-2008,Kov-Kop-Kar-Stu:CQG-2010,Kop-Kar-Kov-Stu:APJ-2010,Kov:EPJP-2013,Kov-Kop-Kar-Koj:CQG-2013}.\footnote{We can alternatively find the equatorial and off-equatorial toroidal structures within a kinetic theory model \cite{Cre-Kov:APJS-2013}; the structures can be also approximated in the framework of strings \cite{Stu-Kol:2012,Tur-Kol:PRD-2013,Ote-Kol-Stu:EPJC-2018}.}

The mentioned model of rotating charged fluid structures is based on an utilization of the energy-momentum conservation and Maxwell equations. Basically, the model is represented by the rotating fluid flow equation -- the momentum conservation equation (general relativistic Euler equation enhanced by a charge term), written in a standard representation; that is with the fluid four-velocity field, pressure, energy and charge density. 
Here, we show that the charged fluid flow can be equivalently described within the momentum conservation equation written in two interesting covariant force representations. The formalism follows from a proper 3+1 spacetime foliation, especially from a projection of the energy-momentum conservation into a proper space-like hypersurface of a given spacetime (hereafter, hypersurface), where forces can be defined.

The first force representation benefits from the hypersurface projection of the fluid four-acceleration field, and from its decomposition in the framework of the so-called 
`optical reference geometry' that was introduced along with the mapping of test particle motion in the proper conformal hypersurface \cite{Abr-Car-Las:GRG-1988,Abr-Pra:MNRAS-1990,Abr-Nur-Wex:CQG-1993,Iye-Pra:CQG-1993,Abr-Nur-Wex:CQG-1995,Son-Mas:CQG-1996,Abr-Las:CQG-1997,Son-Abr:JMP-1998,Stu-Hle:CQG-1999,Stu-Hle-Jur:CQG-2000,Jon:CQG-2006,Jon-Wes:CQG-2006,Son-Abr:JMP-2006,Kov-Stu:IJMPA-2006,Kov-Stu:CQG-2007}. This decomposition enables us to introduce the so-called `inertial forces' that are represented by the velocity and charge independent gravitational force, the velocity dependent and charge independent centrifugal, Coriolis and Euler forces, whereas the sum of these forces vanishes along geodesics -- trajectories of free test particles. Moreover, when a test particle is electromagnetically driven, we get also the so-called `real forces', being represented by the velocity and charge dependent electric and magnetic Lorentz forces \cite{Agu-etal:CQG-1996,Nay-Vis:CQG-1996,Son-Abr:JMP-2006}. Such a force formalism turns out to be very effective, especially for an investigation of test particles motion along circular trajectories \cite{Kov-Stu-Kar:CQG-2008,Kov-Kop-Kar-Stu:CQG-2010,Tur-etal:PRD-2016}.      

The second force representation uniquely follows the first one, enhancing it by a specific enthalpy profile of the fluid, whereas also this representation is related to a special conformal hypersurface geometry. Here, the parental conformal spacetime geometry becomes fairly characteristic as well.  

Concerning conformal spacetimes, in general, special classes of world-lines or trajectories from an ordinary (usually physical) spacetime or hypersurface can be mapped by geodesics of a proper conformal (usually unphysical) spacetime or hypersurface. Consequently, the conformal spacetimes or hypersurfaces become useful for illustration of particular properties of the ordinary ones; this can be effectively visualized by their embedding diagrams.  

The paper is of a theoretical character intended to present the recently introduced model of rotating charged fluid structures from a new alternative perspective providing an interesting mathematical contextualization. In Sec.~2, we describe the charged fluid model in its standard representation. It was introduced in our previous works \cite{Kov-etal:PRD-2011,Kov-etal:PRD-2014,Kov-etal:PRD-2016,Tro-etal:PRD-2018,Schr-etal:PRD-2018,Stu-etal:Uni-2020,Tro-etal:PRD-2020}. Here, however, we include additional  steps of the model development that are necessary to follow new investigation. Firstly, it is the formulation of the charged fluid model in terms of two force representations, as shown in Sec.~3; along with new results presented in Subsecs.~3.1 and 3.2, we remind a particular definition of the forces \cite{Abr-Las:CQG-1997}, which we apply and modify. Secondly, it is the mapping of fluid flow in the related conformal spacetimes and their hypersurfaces, given in Sec.~4. Even if the results presented in Subsec.~4.1 are mostly known from mapping of the test particle motion in conformal hypersurfaces \cite{Abr-Car-Las:GRG-1988,Stu-Hle:CQG-1999,Stu-Hle-Jur:CQG-2000,Son-Abr:JMP-2006,Kov-Stu:CQG-2007}, they enclose and contextualize the brand-new results introduced in Subsec.~4.2. 
Finally, all the obtained essential results are illustrated in Sec.~5, where we construct a rotating charged fluid toroidal structure settled in the equatorial plane of the Schwarzschild spacetime accompanied by an asymptotically uniform magnetic field. Such a configuration was also presented in our previous works \cite{Kov-etal:PRD-2014,Stu-etal:Uni-2020,Tro-etal:PRD-2020}, moreover, with an additional background electric field. Here, however, we choose this configuration to demonstrate the newly introduced force representations of the model, and the related mathematical formalism. Conclusions can be found in Sec.~6.
Hereafter, we use the geometric system of units, $c=G=k_{\rm B}=\nicefrac{1}{4\pi\varepsilon_0}=1$. Moreover, considering particular background fields for a basic demonstration of achieved results, we also scale all the quantities by spacetime mass parameter, $M$; thus, we use the dimensionless units there.
\section{Standard representation of the model}
The presented model of rotating charged fluid structures \cite{Kov-etal:PRD-2011,Kov-etal:PRD-2014,Kov-etal:PRD-2016,Tro-etal:PRD-2018,Schr-etal:PRD-2018,Stu-etal:Uni-2020,Tro-etal:PRD-2020} presumes an existence of gravitational and electromagnetic fields, described by the related spacetime geometry through its metric tensor field, $g_{\alpha\beta}$ (hereafter, metric), and by the Faraday tensor field, $F_{\alpha\beta}$, within the considered Boyer-Linquist coordinates, $x^{\alpha}=(t,r,\theta,\phi)$. As the fundamental assumptions, these background fields must embody the axial symmetry and stationarity, and the fluid is conditioned to be the gravitationally test one, i.e. sufficiently low-mass and weakly-charged not to influence the background gravitational field. 

Neglecting the viscosity and heat conduction, a flow of the considered charged fluid with the locally measured pressure, $p$, and energy density, $\epsilon$, can be mapped by the four-velocity field, $U^{\alpha}$, in terms of the energy-momentum tensor
\begin{eqnarray}
\label{T}
T^{\alpha\beta}&=&(\epsilon+p)U^{\alpha}U^{\beta}+pg^{\alpha\beta}+\frac{1}{4\pi}\left(\mathcal{F}^{\alpha}_{\;\;\gamma}\mathcal{F}^{\beta\gamma}-\frac{1}{4}\mathcal{F}_{\gamma\delta}\mathcal{F}^{\gamma\delta}g^{\alpha\beta}\right).
\end{eqnarray}
Here, $\mathcal{F}^{\alpha\beta}=F^{\alpha\beta}+F^{\alpha\beta}_{\rm self}$, where the background and self-electromagnetic fields satisfy Maxwell equations
\begin{eqnarray}
\label{Maxw1}
\nabla_{\beta}F^{\alpha\beta}=0,\quad
\label{Maxw}
\nabla_{\beta}F^{\alpha\beta}_{\rm self}=4\pi J^{\alpha},
\end{eqnarray}
with $J^{\alpha}$ being the four-current density field of the charged fluid.
Thus, the energy-momentum conservation, $\nabla_{\beta}T^{\alpha\beta}=0$, enables us to write the momentum conservation equation (Euler equation) for the charged fluid flow in the form  
\begin{eqnarray}
\label{Euler}
\partial_{\alpha}p+U_{\alpha}U^{\beta}\partial_{\beta}p\,+(p+\epsilon)U^{\beta}\nabla_{\beta}U_{\alpha}=-J^{\beta}\mathcal{F}_{\beta\alpha}.
\end{eqnarray}

Our model of charged fluid structures is characteristic for its two basic features: elementary charges in the fluid are adherent to the fluid elements, and these are uniformly rotating in the azimuthal direction. Thus, we are interested in the fluid flow mapped by the four-velocity field $U^{\alpha}=(U^t,0,0,U^{\phi})$, where $U^{t}=U^{t}(r,\theta)$ and $U^{\phi}=U^{\phi}(r,\theta)$, which fully determines the four-current density field, $J^{\alpha}=\rho_q U^{\alpha}$. Here, $J^{\phi}$ is the azimuthal flow of the charge density, $\rho_q=q\rho$, where $q$ and $\rho$ are the specific charge and the rest-mass density distributions, respectively.\footnote{The electric current in our model is only of the convective character.}   
Assuming the generated electromagnetic self-field to be much weaker than the background one, $F^{\alpha\beta}_{\rm self}\ll F^{\alpha\beta}$, and the axial symmetry and stationarity of the background fields $g_{\alpha\beta}$ and $F_{\alpha\beta}=\partial_{\alpha}A_{\beta}-\partial_{\beta}A_{\alpha}$, the Euler equation (\ref{Euler}) reduces to the form
\begin{eqnarray}
\label{Euler1}
\partial_{\alpha}p+U_{\alpha}U^{\beta}\partial_{\beta}p\,+(p+\epsilon)U^{\beta}\nabla_{\beta}U_{\alpha}=-\rho_{q}U^{\beta}F_{\beta\alpha},
\end{eqnarray}
and we get the expected stationary and axially symmetric pressure profile of a structure, $p=p(r,\theta)$, as a solution of the
resulting system of `pressure equations' 
\begin{eqnarray}
\label{pressure1}
\partial_{r} p&=&-(p+\epsilon)\Big{(}\partial_{r}\,\ln{|U_t|} - \frac{\Omega \partial_{r} \ell}{1-\Omega \ell}\Big{)}+\rho_q(U^t\partial_{r}  A_t+U^{\phi}\partial_{r}A_{\phi}),\nonumber\\
\partial_{\theta} p&=&-(p+\epsilon)\Big{(}\partial_{\theta}\,\ln{|U_t|} - \frac{\Omega \partial_{\theta} \ell}{1-\Omega \ell}\Big{)}+\rho_q(U^t\partial_{\theta}  A_t+U^{\phi}\partial_{\theta}A_{\phi}).
\end{eqnarray}
Here, the specific angular momentum profile of the fluid, $\ell=-U_{\phi}/U_t$, and the coordinate angular velocity, $\Omega=U^{\phi}/U^t$, are related by the formulae
\begin{eqnarray}
\label{OmegaUt}
\Omega=-\frac{\ell g_{tt}+g_{t\phi}}{\ell g_{t\phi}+g_{\phi\phi}},\quad 
\label{Ut}
U_t=-\Big{(}\frac{g_{t\phi}^2-g_{tt}g_{\phi\phi}}{\ell^2 g_{tt}+2\ell g_{t\phi}+g_{\phi\phi}}\Big{)}^{1/2},
\end{eqnarray}
and the vector potential satisfies the Lorenz gauge, $\nabla_{\alpha}A^{\alpha}=0$. Note that for $\rho_q=0$, the last term in equation (\ref{Euler1}) vanishes, and we get the Euler equation describing a rotating neutral perfect fluid \cite{Koz-Jar-Abr:AA-1978,Abr-Jar-Sik:AA-1978}.

For the given fields $g_{\alpha\beta}$ and $A_{\alpha}$, a solution $p$ of the pressure equations (\ref{pressure1}) can be obtained if an equation of state, $p=p(\rho;q)$ and $\epsilon=\epsilon(p,\rho;q)$, is given, and the rotational regime $\Omega$ and the specific charge distribution $q$, chosen. All these must, however, satisfy an integrability condition for the system of equations (\ref{pressure1}).   

\subsection{Solution in integral form}
Introducing the charge density transformation function   
\begin{eqnarray}
\label{K1}
\mathcal{K}=\frac{\rho_q}{\epsilon+p} U^{\phi},
\end{eqnarray}
and the pressure and electromagnetic vector potential time component transformation relations
\begin{eqnarray}
\label{transf0}
\partial_{\alpha} w=\frac{\partial_{\alpha} p}{(p+\epsilon)},\quad
\partial_{\alpha} a_t = \frac{\partial_{\alpha} A_t}{\Omega},
\end{eqnarray}
where $\alpha$ stands for $r$ and $\theta$, we get the system of transformed pressure equations 
\begin{eqnarray}
\label{pressure4}
\partial_r w=-\partial_r\,\ln{|U_t|} + \frac{\Omega \partial_r \ell}{1-\Omega \ell}+\mathcal{K} \partial_r A\equiv \mathbb{R},\nonumber\\
\partial_{\theta} w=-\partial_{\theta}\,\ln{|U_t|} + \frac{\Omega \partial_{\theta} \ell}{1-\Omega \ell}+\mathcal{K}\partial_{\theta} A\equiv \mathbb{T},
\end{eqnarray} 
where $\partial_{\alpha}A=\partial_{\alpha}a_t+\partial_{\alpha}A_{\phi}$. 
Providing a barotropic fluid, $\epsilon=\epsilon(p)$, and the profiles 
\begin{eqnarray}
\label{generic1}
\Omega=\Omega_{1}(A_t),\quad \Omega=\Omega_{2}(\ell),\quad \mathcal{K}=\mathcal{K}(A),
\end{eqnarray}
ensuring the integrability of equations (\ref{pressure4}) according to the integrability condition 
\begin{eqnarray}
\label{icondition}
\partial_{\theta} \mathbb{R}=\partial_r \mathbb{T},
\end{eqnarray}
we can unify the system of equations (\ref{pressure4}) into the integrable differential form with the solution  
\begin{eqnarray}
\label{united}
w=-\ln{|U_t|}+\ln{|U_{t_{\rm ed}}|}+\int_{\ell_{\rm ed}}^{\ell}\hspace{-0.1cm}\frac{\Omega{\rm d}\ell}{1-\Omega \ell}+\int_{A_{\rm ed}}^{A}\hspace{-0.1cm}\mathcal{K}{\rm d}A\equiv -W+W_{\rm ed},
\end{eqnarray}
where $w=\int_0^w {\rm d}w$.\footnote{Note that the profile $\Omega=\Omega_1(A_t)$ is considered for the integrability of equations (\ref{pressure4}) so that the right hand set of equations (\ref{transf0}) yields $a_t=\int\Omega^{-1}{\rm d}A_t$. Then, $A=a_t+A_{\phi}$, which is necessary for the specification of $\mathcal{K}=\mathcal{K}(A)$.} Here, the function $W(r,\theta)$ stands for the potential (variable part of $w$), and the subscript `${\rm ed}$' relates to the position of an edge of the structure at $r=r_{\rm ed}$ and $\theta=\theta_{\rm ed}$, determining constants of integration, all being coupled in $W_{\rm ed}$. Thanks to the transformation (\ref{transf0}), being integrated as $w = \int_0^p \frac{{\rm d}p}{p+\epsilon}$, the equipressure surfaces, $p={\rm const}$, determining the shape of a possible fluid structure, are of the same form as the equipotential surfaces, $W={\rm const}$. Moreover, in the case of the isentropic flow of the fluid, ${\rm d}h-{\rm d}p/\rho=0$, the function $w$ is related directly to the specific enthalpy of the fluid, $h$, by the formula  
\begin{eqnarray}
\label{enthalpy}
w=\ln{h},\quad h=\frac{\epsilon+p}{\rho}.
\end{eqnarray}

Formally, the solution (\ref{united}) follows from an integration of the Pfaffian form 
\begin{eqnarray}
{\rm d}w(\ln{U_t},\ell, A)\equiv -{\rm d}\ln{|U_t|}+\frac{\Omega{\rm d}\ell}{1-\Omega \ell}+\mathcal{K}{\rm d}A.
\end{eqnarray}
Thus, within the standard representation of the model, we analytically describe the fluid structures in terms of solutions determined by the Pfaffian coordinates $(\ln{|U_t|},\ell,A)$ and parametrized by the generic functions -- the restricting conditions (\ref{generic1}).\footnote{Particular profiles of the generic functions and the consideration of barotropic fluid, all required for the integrability of equations (\ref{pressure1}) providing an analytic solution, match wide range of physical scenarios; however, they restrict degrees of freedom in the model.}

Finally, getting back to the important characteristic of our model -- the charged density profile $\rho_q=q\rho$, particularly to the specific charge profile $q$, we find it directly determined by the chosen profiles of $\mathcal{K}$ and $\Omega$ (\ref{generic1}), consequently also by the related solution $w$, as it is clear from relations (\ref{OmegaUt}) and $q=\mathcal{K}{\rm e}^w/U^{\phi}$, following from (\ref{K1}) and (\ref{enthalpy}).
 
\section{Force formalism}
Within general relativity, a generally covariant definition of forces follows from a suitable 3+1 foliation of a given spacetime into space-like hypersurfaces orthogonal to a particular congruence of time-like world-lines. Here, we choose the congruence corresponding to the world-lines of a class of stationary observers characterized by a time-like unit four-velocity field, $n^{\alpha}$, satisfying the conditions \cite{Abr-Las:CQG-1997}
\begin{eqnarray}
\label{1}
n^{\alpha} n_{\alpha}=-1,\quad
n^{\beta}\nabla_{\beta} n_{\alpha}=h^{\beta}_{\;\;\alpha}\partial_{\beta}\Phi, \quad n_{[\alpha}\nabla_{\beta} n_{\gamma]}=0.
\end{eqnarray}
Thus, along with the normalization condition and the condition of hypersurface orthogonality, we require the four-acceleration field of the observers, $n^{\beta}\nabla_{\beta} n_{\alpha}$, to be formulable in terms of an unspecified scalar function $\Phi$; the form of $\Phi$ depends on a particular choice of $n^{\alpha}$.

Then, the four-velocity field of the fluid elements $U^{\alpha}$ can be decomposed into parts parallel and orthogonal to the four-vector field $n^{\alpha}$  
\begin{eqnarray}
\label{8}
U^{\alpha}=\gamma(n^{\alpha}+v\tau^{\alpha}),
\end{eqnarray}
where $\tau^{\alpha}$ is a space-like unit four-vector field. Together with the velocity field of the fluid elements $v$ with respect to the chosen world-line congruence,  the vector field  $\tau^{\alpha}$ maps the fluid flow in the hypersurface -- determines the fluid flow lines; it arises from the four-velocity field hypersurface projection  
\begin{eqnarray}
h^{\alpha}_{\;\;\beta}U^{\beta}=\gamma v\tau^{\alpha},\quad \gamma=-n^{\alpha} U_{\alpha},\quad \gamma=(1-v^2)^{-1/2},
\end{eqnarray}
with the projection tensor   
\begin{eqnarray}
\label{proj}
h^{\alpha}_{\;\;\beta}=\delta^{\alpha}_{\;\;\beta}+n^{\alpha} n_{\beta}.
\end{eqnarray}

The hypersurface projection of the four-acceleration formula can be decomposed according to the orders of the velocity $v$ into four parts \cite{Abr-Las:CQG-1997}
\begin{eqnarray}
\label{11}
h^{\gamma}_{\;\;\alpha}U^{\beta}\nabla_{\beta}U_{\gamma}=-\tilde{G}_{\alpha}^{\perp}-\tilde{Z}_{\alpha}^{\perp}-\tilde{C}_{\alpha}^{\perp}-\tilde{L}_{\alpha}^{\perp},
\end{eqnarray}
where the four-acceleration terms are expressed as   
\begin{eqnarray}
\label{G}
\tilde{G}_{\alpha}^{\perp}=-h^{\beta}_{\;\;\alpha}\partial_{\beta}\Phi,\\
\label{Z}
\tilde{Z}_{\alpha}^{\perp}=-(\gamma v)^{2}(\tau^{\beta}\nabla_{\beta}\tau_{\alpha}-h^{\beta}_{\;\;\alpha}\partial_{\beta}\Phi-\tau_{\alpha}\tau^{\beta}\partial_{\beta}\Phi),\\
\label{C}
\tilde{C}_{\alpha}^{\perp}=-\gamma^2v n^{\beta}(\nabla_{\beta}\tau_{\alpha}-\nabla_{\alpha}\tau_{\beta}),\\
\label{L}
\tilde{L}_{\alpha}^{\perp}=-U^{\beta}\partial_{\beta}(\gamma v {\rm e}^{\Phi}){\rm e}^{-\Phi}\tau_{\alpha}.
\end{eqnarray}
Note that in some special cases, the decomposition (\ref{11}) admits a formal comparison of terms (\ref{G})--(\ref{L}) with their 3-dimensional Newtonian counterparts, accordingly naming them as `inertial gravitational, centrifugal, Coriolis and Euler accelerations, respectively -- specific forces' (hereafter, forces), and referring to the scalar function $\Phi$ as a `gravitational potential' and its negative gradient, $-\partial_{\alpha}\Phi$, as a `gravitational intensity'. Apart from such a Newtonian terminology, one can appreciate this decomposition as an illustrative tool for surveying particular problems of relativistic dynamics.\footnote{The superscript $^\perp$ emphasizes the outcome of the hypersurface projection. In the following sections, however, it is dropped to distinguish between the terms defined for the general motion defined here, and those restricted to the case of an uniform circular motion introduced later.} 

The `Newtonian-like reading' can be applicable also in our case of the stationary and axially symmetric spacetime with the related time-like, $\eta^{\alpha}=(1,0,0,0)$, and space-like, $\xi^{\alpha}=(0,0,0,1)$, Killing four-vector fields, where we choose the splitting world-line congruence as the congruence of world-lines of zero angular momentum observers. Their four-velocity field can be written as    
\begin{eqnarray}
\label{3}
n^{\alpha}={\rm e}^{-\Phi}\iota^{\alpha},\quad \Phi=\nicefrac{1}{2}\ln{(-\iota^{\beta}\iota_{\beta})},\quad \iota^{\alpha}=\eta^{\alpha}+\Omega^{\rm o}\xi^{\alpha},
\end{eqnarray}
where $\Omega^{\rm o}=-\eta^{\alpha}\xi_{\alpha}/\xi^{\alpha}\xi_{\alpha}=-g_{t\phi}/g_{\phi\phi}$. The parallel four-vector fields $\iota^{\alpha}$ and $n^{\alpha}$ are orthogonal to the space-like hypersurface $\Sigma_t$ ($t={\rm const}$); its 3-dimensional geometry can be described by the metric $h_{\alpha\beta}=g_{\alpha\beta}+n_{\alpha} n_{\beta}$, especially, by its components with $\alpha$ and $\beta$ standing for $r$, $\theta$ and $\phi$, the so-called `induced metric'. The choice of splitting world-line congruence given by the tangent four-vector field (\ref{3}) provides suitably     $h^{\beta}_{\;\;\alpha}\partial_{\beta}\Phi=\partial_{\alpha}\Phi$, allowing us to formally write the gravitational force as negative gradient of the gravitational potential $\tilde{G}_{\alpha}^{\perp}=-\partial_{\alpha}\Phi$.

On the other hand, it is important to stress that the decomposition (\ref{11}) is not fundamentally unique. In more complicated (less symmetric) backgrounds or for other choices of $n^{\alpha}$, its Newtonian reading  fails. The decomposition, however, remains as generally valid formalism in relativistic dynamics \cite{Jon:CQG-2006,Jon-Wes:CQG-2006}, as well as different kinds of four-acceleration decomposition can be employed \cite{Fel:CQG-1991,Fel:MNRAS-1991,Bin-etal:IJMPD-1997a,Bin-etal:IJMPD-1997b}.

\subsection{The $\Phi$-force representation of the model}
\label{Phi-forces}
Assuming the world-line congruence given by the four-vector field (\ref{3}) orthogonal to the related hypersurface $\Sigma_t$, the hypersurface projection of the Euler equation (\ref{Euler1}) characterizing our model can be written in the force representation as 
\begin{eqnarray}
\label{Euler2}
\partial_{\alpha}p-(p+\epsilon)(\tilde{G}_{\alpha}+\tilde{Z}_{\alpha}+\tilde{C}_{\alpha})=\rho(\tilde{E}_{\alpha}+\tilde{M}_{\alpha}),
\end{eqnarray}
where 
\begin{eqnarray}
\label{G1}
\tilde{G}_{\alpha}&=&-\partial_{\alpha}\Phi,\\
\label{Z1}
\tilde{Z}_{\alpha}&=&\frac{\,\tilde{v}^2}{\tilde{R}}\,\partial_{\alpha}\tilde{R},\\
\label{C1}
\tilde{C}_{\alpha}&=&-(1+\tilde{v}^2)^{1/2}\tilde{v}\tilde{R}\,\partial_{\alpha}\Omega^{\rm o},\\
\label{E1}
\tilde{E}_{\alpha}&=&q {\rm e}^{-\Phi}(1+\tilde{v}^2)^{1/2}(\eta^{\beta}+\Omega^{\rm o}\xi^{\beta})\,\partial_{\alpha}A_{\beta},\\
\label{M1}
\tilde{M}_{\alpha}&=&q {\rm e}^{-\Phi} \frac{\tilde{v}}{\tilde{R}}\xi^{\beta}\,\partial_{\alpha}A_{\beta},
\end{eqnarray}
and  
\begin{eqnarray}
\label{41}
\tilde{R}={\rm e}^{-\Phi}g_{\phi\phi}^{1/2},\quad\tilde{v}&=&\gamma v,\quad v=\tilde{R}\tilde\Omega,\quad \tilde{\Omega}=\Omega-\Omega^{\rm o}.  
\end{eqnarray}
The sum of the inertial gravitational, centrifugal and Coriolis forces follows from the projection of the general four-acceleration field of the rotating fluid, $\tilde{G}_{\alpha}+\tilde{Z}_{\alpha}+\tilde{C}_{\alpha}=-h^{\gamma}_{\;\;\alpha} U^{\beta}\nabla_{\beta}U_{\gamma}$, while the sum of the real electric and magnetic forces follows from the projection of the Lorentz term, $\tilde{E}_{\alpha}+\tilde{M}_{\alpha}=-q h^{\gamma}_{\:\:\:\alpha} U^{\beta}F_{\beta\gamma}=q \gamma n^{\beta}\,\partial_{\alpha}A_{\beta}+q \tilde{v} \tau^{\beta}\,\partial_{\alpha}A_{\beta}$. Note that formulae (\ref{G1})-(\ref{C1}) follow directly also from their general forms (\ref{G})-(\ref{C}) being restricted to the case of uniform and axially symmetric fluid rotation
\begin{eqnarray}
\label{circulation}
\tau^{\alpha}=g_{\phi\phi}^{-1/2}\xi^{\alpha}, \quad v=v(r,\theta),
\end{eqnarray}
in accordance with the basic features of the model; such a regime of rotation also means vanishing of the Euler force (\ref{L}). 

The scaled cylindrical radius $\tilde{R}$ defined in terms of the scaling factor ${\rm e}^{-\Phi}$, entitles this force representation as the $\Phi$-force representation, and directly determines behaviour of the centrifugal force (\ref{Z1}), especially of its velocity independent part 
\begin{eqnarray}
\label{vel-centr-1}
\tilde{\mathcal{Z}}_{\alpha}\equiv\frac{\tilde{Z}_{\alpha}}{\tilde{v}^2}=\frac{\partial_{\alpha}\tilde{R}}{\tilde{R}}=-\partial_{\alpha} \ln{\tilde{R}^{-1}}.
\end{eqnarray}
The field $\tilde{\mathcal{Z}}_{\alpha}$ is equal to negative gradient of the scalar function $\ln{\tilde{R}^{-1}}$, which we consider as a `cylindrical potential'; thus, $\tilde{\mathcal{Z}}_{\alpha}$ represents  a `cylindrical intensity'. As well as the gravitational potential $\Phi$, with its negative gradient defining the gravitational intensity $\tilde{G}_{\alpha}$, the potential  $\ln{\tilde{R}^{-1}}$ is of the pure spacetime character; 
the radius $\tilde{R}$ corresponds also to the so-called `radius of gyration' \cite{Abr-Mil-Stu:PRD-1993,Abr-Nur-Wex:CQG-1995}. Note that both the electric (\ref{E1}) and magnetic (\ref{M1}) forces contain the same `electro-gravitational' term $q {\rm e}^{-\Phi}$. In the expression for the electric force $\tilde{E}_{\alpha}$, the factor ${\rm e}^{-\Phi}$ appears `naturally' as the normalization factor from $n^{\alpha}={\rm e}^{-\Phi}\iota^{\alpha}$, while in the expression for the magnetic force $\tilde{M}_{\alpha}$, the factor ${\rm e}^{-\Phi}$ appears `technically' as the scaling factor from $\tilde{R}=g_{\phi\phi}^{-1/2}{\rm e}^{-\Phi}$.

In the case of uniform circular motion of charged test particles, the general Lorentz equation of motion, $U^{\beta}\nabla_{\beta}U_{\alpha}=-qU^{\beta}F_{\beta\alpha}$, reduces to the simple force form 
\begin{eqnarray}
\label{e-geodesic}
-\tilde{G}_{\alpha}-\tilde{Z}_{\alpha}-\tilde{C}_{\alpha}=\tilde{E}_{\alpha}+\tilde{M}_{\alpha},
\end{eqnarray}
whereas the equation $\tilde{G}_{\alpha}+\tilde{Z}_{\alpha}+\tilde{C}_{\alpha}=0$ determines circular geodesics.

The force representation of the Euler equation (\ref{Euler2}) represents the system of pressure equations equivalent to the ones provided by the standard representation (\ref{pressure1}). By using the charge density transformation function (\ref{K1}) and the pressure and electromagnetic vector potential time component transformation relations (\ref{transf0}), we get the system of transformed pressure equations 
\begin{eqnarray}
\label{pressure5}
\partial_{r} w=-\partial_r \Phi + \frac{\,\tilde{v}^2}{\tilde{R}}\,\partial_{r}\tilde{R}-(1+\tilde{v}^2)^{1/2}\tilde{v}\tilde{R}\,\partial_{r}\Omega^{\rm o}+\mathcal{K} \partial_{r} A\equiv \tilde{\mathbb{R}},\nonumber\\
\partial_{\theta} w=-\partial_{\theta}\Phi+\frac{\,\tilde{v}^2}{\tilde{R}}\,\partial_{\theta}\tilde{R}-(1+\tilde{v}^2)^{1/2}\tilde{v}\tilde{R}\,\partial_{\theta}\Omega^{\rm o}+\mathcal{K} \partial_{\theta} A\equiv \tilde{\mathbb{T}}.
\end{eqnarray} 
Providing the profiles 
\begin{eqnarray}
\label{generic2}
\Omega=\Omega(A_t),\quad \mathcal{K}=\mathcal{K}(A),\quad \tilde{v}=\tilde{v}(\tilde{R}),\quad (1+\tilde{v}^2)^{1/2}\tilde{v}\tilde{R}=\mathcal{V}(\Omega^{\rm o}),
\end{eqnarray}
for the barotropic fluid, $\epsilon=\epsilon(p)$, the integrability condition 
\begin{eqnarray}
\label{iicondition}
\partial_{\theta} \tilde{\mathbb{R}}=\partial_r \tilde{\mathbb{T}},
\end{eqnarray}
is satisfied, and we can unify the system of equations (\ref{pressure5}) into the integrable differential form with the solution 
\begin{eqnarray}
\label{united1}
w=-\Phi+\Phi_{\rm ed}+\int_{\tilde{R}_{\rm ed}}^{\tilde{R}} \frac{\,\tilde{v}^2}{\tilde{R}}\,{\rm d}\tilde{R}-\int_{{\Omega^{\rm o}}_{\hspace{-0.1cm} \rm ed}}^{\Omega^{\rm o}} \mathcal{V}{\rm d}\Omega^{\rm o}+\int_{A_{\rm ed}}^{A}\hspace{-0.1cm}\mathcal{K}{\rm d}A\equiv -\tilde{W}+\tilde{W}_{\rm ed}.
\end{eqnarray}
Note that the transformation function defined by relation (\ref{K1}) can be rewritten into the form 
\begin{eqnarray}
\label{K2}
\mathcal{K}=\frac{\rho_q}{\epsilon+p} \gamma {\rm e}^{-\Phi}\Omega,
\end{eqnarray}
since $U^t={\rm e}^{-\Phi}\gamma$; we also remind that $\Omega=\Omega^{\rm o}+v/\tilde{R}$, as it follows from relation (\ref{41}). 

Within the $\Phi$-force representation of the model, the solution (\ref{united1}) follows from an integration of the Pfaffian form 
\begin{eqnarray}
{\rm d}w(\Phi,\tilde{R},\Omega^{\rm o},A)\equiv-{\rm d}\Phi+\frac{\,\tilde{v}^2}{\tilde{R}}\,{\rm d}\tilde{R}-\mathcal{V}\,{\rm d}\Omega^{\rm o}+\mathcal{K}{\rm d}A,
\end{eqnarray}
determined by different Pfaffian coordinates, in comparison with the solution (\ref{united}). Since the number of coordinates and related restricting conditions (\ref{generic2}) is greater here, searching for an analytical solution can be less general. On the other hand, we can clearly identify the pure spacetime frame dragging part  of the solution, embodied in the Coriolis term associated with the Pfaffian coordinate $\Omega^{\rm o}$. The representation is very intuitive in static spacetimes. For instance, there the Newtonian limit of the solution (\ref{united1}) can be easily found, containing the quantities having their direct Newtonian counterparts $\Phi \rightarrow \Phi_{\rm N}$, $\tilde{R}\rightarrow R_{\rm N}$, $\tilde{v}\rightarrow v_{\rm N}$. 

\subsection{The $w$-force representation of the model}
By using relations (\ref{transf0}) and (\ref{enthalpy}), the $\Phi$-force representation (\ref{Euler2}) of the Euler equation (\ref{Euler1}) can be rewritten into the  equivalent form 
\begin{eqnarray}
\label{Euler3}
-\bar{G}_{\alpha}-\bar{Z}_{\alpha}-\bar{C}_{\alpha}=\bar{E}_{\alpha}+\bar{M}_{\alpha},
\end{eqnarray}
where 
\begin{eqnarray}
\label{G2}
\bar{G}_{\alpha}&=&-\partial_{\alpha}\Psi,\\
\label{Z2}
\bar{Z}_{\alpha}&=&v^2\,\frac{\partial_{\alpha}\bar{R}}{\bar{R}},\\
\label{C2}
\bar{C}_{\alpha}&=&-v{\rm e}^{-\Phi}\bar{R}\,\partial_{\alpha}\Omega^{\rm o},\\
\bar{E}_{\alpha}&=&q\frac{{\rm e}^{-\Psi}}{\gamma^{2}}(\eta^{\beta}+\Omega^{\rm o}\xi^{\beta})\,\partial_{\alpha}A_{\beta},\\
\bar{M}_{\alpha}&=&q\frac{v}{\gamma^{2}\bar{R}}\xi^{\beta}\,\partial_{\alpha}A_{\beta},
\end{eqnarray}
and  
\begin{eqnarray}
\label{Psi-pot}
\Psi=\Phi+w,\quad \bar{R}={\rm e}^w g_{\phi\phi}^{1/2}={\rm e}^{\Psi}\tilde{R}.
\end{eqnarray}
We can see that the inertial gravitational, $\bar{G}_{\alpha}$, centrifugal, $\bar{Z}_{\alpha}$, and Coriolis, $\bar{C}_{\alpha}$, forces, and the real electric, $\bar{E}_{\alpha}$, and magnetic, $\bar{M}_{\alpha}$, forces defined here resemble the forces introduced in Subsec.~\ref{Phi-forces}. Moreover, the force equation (\ref{Euler3}) formally coincides with the one written for the charged test particles (\ref{e-geodesic}).    

However, here the gravitational potential $\Psi$ is represented by the function containing a contribution from the specific enthalpy through the term $w=\ln{h}$. Thus, in comparison with the gravitational potential $\Phi$ of the representation (\ref{Euler2}), this one is not of the pure geometry nature, but depends also on the fluid flow, possibly forming a structure. This is obvious also for the scaled cylindrical radius, $\bar{R}$, which is defined in terms of the specific enthalpy, $h={\rm e}^w$, directly. Such a definition entitles this force representation as the $w$-force representation, where the radius $\bar{R}$ directly determines behaviour of the centrifugal force (\ref{Z2}), especially of its velocity independent part
\begin{eqnarray}
\label{vel-centr-2}
\bar{\mathcal{Z}}_{\alpha}\equiv\frac{\bar{Z}_{\alpha}}{v^2}=\frac{\partial_{\alpha}\bar{R}}{\bar{R}}=-\partial_{\alpha} \ln{\bar{R}^{-1}}.
\end{eqnarray}
Also here, the field $\bar{\mathcal{Z}}_{\alpha}$ is equal to negative gradient of the scalar function $\ln{\bar{R}^{-1}}$, which we consider as a cylindrical potential. Thus, $\bar{\mathcal{Z}}_{\alpha}$ represents a cylindrical intensity; the gravitational intensity corresponds to the gravitational force $\bar{G}_{\alpha}$.
\section{Conformal geometries and embedding diagrams}
By rescaling a metric $g_{\alpha\beta}$ of an ordinary spacetime $\mathcal{S}$ geometry by a proper smooth scalar function, $\mathcal{C}^2$, we get the conformal metric,  $\tilde{g}_{\alpha\beta}=\mathcal{C}^2 g_{\alpha\beta}$, describing the conformal spacetime $\tilde{\mathcal{S}}$ geometry. In general, unlike angles, norms of time-like and space-like four-vectors change under this conformal transformation. And, except null geodesics, geodesics from $\mathcal{S}$ transform to (otherwise, are mapped by) general world-lines in $\tilde{\mathcal{S}}$. On the other hand, general world-lines from $\mathcal{S}$ can be mapped by geodesics of $\tilde{\mathcal{S}}$, otherwise also by another unique class of world-lines.
The same holds for the induced metric $h_{\alpha\beta}$ of an ordinary hypersurface $\Sigma\subset\mathcal{S}$ geometry, for the induced conformal metric, $\tilde{h}_{\alpha\beta}=\mathcal{C}^2 h_{\alpha\beta}$, of the conformal hypersurface $\tilde{\Sigma}\subset\tilde{\mathcal{S}}$ geometry, and for mapping of geodesics and general trajectories from $\Sigma$ by general trajectories in $\tilde{\Sigma}$, and vice versa.  
Consequently, a specific feature of an ordinary spacetime, reflected in the general relativistic dynamics, can be effectively illustrated by a proper conformal spacetime geometry, or even visualized in terms of its embedding diagrams. 

An embedding diagram of a given stationary and axially symmetric spacetime $\mathcal{S}$ represents mapping of its two-dimensional surface into the three-dimensional Euclidean space $\mathcal{E}$. Here, having the geometry described by the cylindrical coordinates, $(R,\alpha,z)$, and by the related line element, $d\sigma_{\mathcal{E}}^2={\rm d}R^2+R^2{\rm d}\alpha^2+{\rm d}z^2$, we can assume a rotational, $z=z(R)$, surface determined by the line element 
\begin{eqnarray}
\label{2e}
{\rm d}l_{\rm \mathcal{E}}^2=\Big{[}1+{\Big{(}\frac{{\rm d}z}{{\rm d}R}\Big{)}}^2\Big{]}{\rm d}R^2+R^2{\rm d}\alpha^2,
\end{eqnarray}
and consider it to be isometric -- usually to the equatorial plane, $\theta=\nicefrac{\pi}{2}$, of the hypersurface $\Sigma_t$; this can be described by the line element 
\begin{eqnarray}
\label{3e}
{\rm d}l^2=h_{rr}{\rm d}r^2+h_{\phi\phi}{\rm d}\phi^2, 
\end{eqnarray}
where $h_{rr}$ and $h_{\phi\phi}$ are the induced metric coefficients of the $\Sigma_t$ geometry. Note that in this case, we speak of the embedding diagram of equatorial plane of the hypersurface $\Sigma_t$ (hereafter, embedding diagram of $\Sigma_t$ equatorial plane).
Assigning the coordinates so that $\alpha=\phi$ and $R^2=h_{\phi\phi}$, we obtain the embedding formula, $z=z(R)$, or its equivalent, $z=z(r)$, written in the differential forms 
\begin{eqnarray}
\label{emb0}
\Big{(}\frac{{\rm d}z}{{\rm d}R}\Big{)}^2=h_{rr}\Big{(}\frac{{\rm d}r}{{\rm d}R}\Big{)}^2-1,\quad 
\label{emb}
\frac{{\rm d}z}{{\rm d}r}=\pm\Big{[}h_{rr}-\Big{(}\frac{{\rm d}R}{{\rm d}r}\Big{)}^2\Big{]}^{1/2},
\end{eqnarray}
where ${\rm d}R/{\rm d}r={\rm d}\sqrt{h_{\phi\phi}}/{\rm d}r$; the signs in formula lead to isometric surfaces \cite{Stu-Hle-Jur:CQG-2000}. 

In general, shape of an embedding diagram of equatorial plane is characterized by its steepness and turning points, which are determined by the relation ${\rm d}R/{\rm d}z=0$. Equivalently, because of the relation
\begin{eqnarray}
\frac{{\rm d}z}{{\rm d}R}=\frac{{\rm d}z}{{\rm d}r}\frac{{\rm d}r}{{\rm d}R},
\end{eqnarray}
the turning points are also determined by the condition ${\rm d}R/{\rm d}r=0$.
\subsection{Conformal transformation $\mathcal{C}={\rm e}^{-\Phi}$}
Geometry of a conformal spacetime $\tilde{\mathcal{S}}$ described by the metric 
\begin{eqnarray}
\label{metric1}
\tilde{g}_{\alpha\beta}={\rm e}^{-2\Phi}g_{\alpha\beta},
\end{eqnarray}
is well known for its conformal hypersurface $\tilde{\Sigma_t}\subset\tilde{\mathcal{S}}$ geometry, the so-called `optical reference geometry' \cite{Abr-Car-Las:GRG-1988,Abr-Pra:MNRAS-1990,Abr-Nur-Wex:CQG-1993,Iye-Pra:CQG-1993,Abr-Nur-Wex:CQG-1995,Son-Mas:CQG-1996,Abr-Las:CQG-1997,Son-Abr:JMP-1998,Stu-Hle:CQG-1999,Stu-Hle-Jur:CQG-2000,Jon:CQG-2006,Jon-Wes:CQG-2006,Son-Abr:JMP-2006,Kov-Stu:IJMPA-2006,Kov-Stu:CQG-2007}, being described by the metric 
\begin{eqnarray}
\tilde{h}_{\alpha\beta}={\rm e}^{-2\Phi}(g_{\alpha\beta}+n_{\alpha} n_{\beta}).
\end{eqnarray}  

The main feature of this conformal hypersurface $\tilde{\Sigma}_t$ is the correspondence between its geodesics determined by the relation
\begin{eqnarray}
\label{geodesics1}
\tilde{\tau}^{\alpha}\tilde{\nabla}_{\alpha}\tilde{\tau}_{\beta}=\tau^{\alpha}\nabla_{\alpha}\tau_{\beta}-\partial_{\beta}\Phi-\tau_{\beta}\tau^{\alpha}\partial_{\alpha}\Phi=0,
\end{eqnarray} 
where $\tilde{\nabla}_{\alpha}$ is the covariant derivative with respect to the metric $\tilde{h}_{\alpha\beta}$, $n^{\alpha}$ is given by relations (\ref{3}), $\tilde{\tau}^{\alpha}={\rm e}^{\Phi}\tau^{\alpha}$,  $\tilde{\tau}_{\alpha}={\rm e}^{-\Phi}\tau_{\alpha}$, and trajectories of light in the ordinary hypersurface $\Sigma_t$; this is, however, true only for static ordinary spacetimes $\mathcal{S}$. 
In general, geodesics of $\tilde{\Sigma}_t$ correspond to trajectories from $\Sigma_t$ determined by the relation $\tilde{\mathcal{Z}}^{\perp}_{\alpha}=0$. This is because the centrifugal force takes the form
\begin{eqnarray}
\label{centr1}
\tilde{Z}_{\alpha}^{\perp}&=&-\tilde{v}^2\tilde{\tau}^{\beta}\tilde{\nabla}_{\beta}\tilde{\tau}_{\alpha},
\end{eqnarray}
in the case of a general motion, as it follows from relations (\ref{Z}) and (\ref{geodesics1}) \cite{Abr-Car-Las:GRG-1988,Abr-Pra:MNRAS-1990,Abr-Nur-Wex:CQG-1993,Iye-Pra:CQG-1993,Abr-Nur-Wex:CQG-1995,Son-Mas:CQG-1996,Abr-Las:CQG-1997,Son-Abr:JMP-1998,Stu-Hle:CQG-1999,Stu-Hle-Jur:CQG-2000,Jon:CQG-2006,Jon-Wes:CQG-2006,Son-Abr:JMP-2006,Kov-Stu:IJMPA-2006,Kov-Stu:CQG-2007}. Thus, the fluid flow lines in $\Sigma_t$ along which the centrifugal force vanishes independently of velocity are mapped by geodesics in $\tilde{\Sigma}_t$.  
For the uniform and axially symmetric rotation (\ref{circulation}), the centrifugal force formula (\ref{centr1}) reduces to formula (\ref{Z1}). 

The embedding diagram of $\tilde{\Sigma}_t$ equatorial plane is governed by the general embedding formula (\ref{emb}). Here, taking its particular form
\begin{eqnarray}
\label{emb1}
\frac{{\rm d}z}{{\rm d}r}=\pm\Big{[}\tilde{h}_{rr}-\Big{(}\frac{{\rm d}\tilde{R}}{{\rm d}r}\Big{)}^2\Big{]}^{1/2},
\end{eqnarray}
the formula provides turning points of the diagram determined by the condition ${\rm d}\tilde{R}/{\rm d}r=0$, i.e. by the condition $\tilde{\mathcal{Z}}_r=0$ (\ref{vel-centr-1}) \cite{Abr-Car-Las:GRG-1988,Stu-Hle:CQG-1999,Stu-Hle-Jur:CQG-2000,Son-Abr:JMP-2006,Kov-Stu:CQG-2007}. Thus, in the case of a fluid rotation, equatorial flow lines along which the radial component of cylindrical intensity vanishes show up as turning circles of throats or bellies of the embedding diagram.

\subsection{Conformal transformation $\mathcal{C}={\rm e}^{w}$}
Unlike the previous conformal rescaling (\ref{metric1}), geometry of a conformal spacetime $\bar{\mathcal{S}}$ described by the metric
\begin{eqnarray}
\label{metric2}
\bar{g}_{\alpha\beta}={\rm e}^{2w}g_{\alpha\beta},
\end{eqnarray}
where $h={\rm e}^{w}$ is the specific enthalpy (\ref{enthalpy}), is based on the scaling factor which is not of the pure $\mathcal{S}$ geometry character. As such it depends on the solution describing flow of the fluid, possibly forming a structure. Note that the type of metric (\ref{metric2}) has been already mentioned in the literature \cite{Ani:1989}, but related only to a neutral fluid flow; here, we introduce its generalization to the case of charged fluid. 

In the case of isentropic fluid flow, the neutral fluid flow world-lines from the ordinary spacetime $\mathcal{S}$  are mapped by geodesics of $\bar{\mathcal{S}}$ determined by the relation
\begin{eqnarray}
\bar{U}^{\beta}\bar{\nabla}_{\beta}\bar{U}_{\alpha}=U^{\beta}\nabla_{\beta}U_{\alpha}+\partial_{\alpha}w+U_{\alpha}U^{\beta}\partial_{\beta}w=0,
\end{eqnarray} 
where $\bar{\nabla}_{\alpha}$ is the covariant derivative with respect to the metric $\bar{g}_{\alpha\beta}$, $\bar{U}^{\alpha}={\rm e}^{-w}U^{\alpha}$, $\bar{U}_{\alpha}={\rm e}^{w}U_{\alpha}$ \cite{Ani:1989}. In the case of charged fluid, the fluid flow world-lines from $\mathcal{S}$ are mapped by world-lines determined by the `Lorentz-like' equation in $\bar{\mathcal{S}}$. Namely, the Euler equation (\ref{Euler}) can be written in the form
\begin{eqnarray}
\label{conformal-Lorentz}
\bar{U}^{\beta}\bar{\nabla}_{\beta}\bar{U}_{\alpha}=-q\bar{U}_{\beta}F^{\beta}_{\;\;\alpha},
\end{eqnarray}
where the fluid flow is supposed to be characterized only by the convective transport of charge, $\sigma=0$, as it is considered in our model of the rotating charged fluids. Particularly, we obtain the conformal equivalent of the pressure equations (\ref{pressure1}) written in the form   
\begin{eqnarray}
\label{pressure2}
-\partial_r\,\ln{|\bar{U}_t|} + \frac{\Omega \partial_r \ell}{1-\Omega \ell}+q(\bar{U}^t\partial_r A_t+\bar{U}^{\phi}\partial_r A_{\phi})=0,\nonumber\\
-\partial_{\theta}\,\ln{|\bar{U}_t|} + \frac{\Omega \partial_{\theta} \ell}{1-\Omega \ell}+q(\bar{U}^t\partial_{\theta} A_t+\bar{U}^{\phi}\partial_{\theta} A_{\phi})=0.
\end{eqnarray}

Note that since the charged fluid dynamics described by the Lorentz-like equation (\ref{conformal-Lorentz}) contains the specific charge $q$ that is, in general not constant, the dynamics of charged test particles in fixed gravitational and electromagnetic field is modified in the fluid description. We qualitatively understand the dynamics modified by the fluid effects by equation (\ref{pressure2}), in which a ratio of charge and mass densities plays a key role.
The specific charge of fluid element, $q=\rho_q/\rho$, in $\mathcal{S}$ represents also the specific charge characterizing the world-line in $\bar{S}$.
 
The main feature of the conformal hypersurface $\bar{\Sigma}_t$ with its geometry described by the metric  
\begin{eqnarray}
\bar{h}_{\alpha\beta}={\rm e}^{2w}(g_{\alpha\beta}+n_{\alpha} n_{\beta}),
\end{eqnarray}  
is the correspondence between its geodesics determined by the relation
\begin{eqnarray}
\bar{\tau}^{\alpha}\bar{\nabla}_{\alpha}\bar{\tau}_{\beta}=\tau^{\alpha}\nabla_{\alpha}\tau_{\beta}+\partial_{\beta}w+\tau_{\beta}\tau^{\alpha}\partial_{\alpha}w=0,
\end{eqnarray} 
where  $\bar{\tau}^{\alpha}={\rm e}^{-w}\tau^{\alpha}$, $n^{\alpha}$ is given by relations (\ref{3}), $\tilde{\tau}_{\alpha}={\rm e}^{w}\tau_{\alpha}$, and trajectories from $\Sigma_t$ determined by the relation $\bar{\mathcal{Z}}^{\perp}_{\alpha}=0$. If the centrifugal force is defined as    
\begin{eqnarray}
\label{centr2}
\bar{Z}_{\alpha}^{\perp}&=&-v^2\bar{\tau}^{\beta}\bar{\nabla}_{\beta}\bar{\tau}_{\alpha},
\end{eqnarray}
in the case of general motion, fluid flow lines in $\Sigma_t$ along which the centrifugal force vanishes independently of velocity are mapped by geodesics in $\bar{\Sigma}_t$. For the uniform and axially symmetric rotation (\ref{circulation}), the centrifugal force formula (\ref{centr2}) reduces to formula (\ref{Z2}).

Constructing the embedding diagram of $\bar{\Sigma}_t$ equatorial plane, we can see that the general embedding formula (\ref{emb}) takes the particular form
\begin{eqnarray}
\label{emb2}
\frac{{\rm d}z}{{\rm d}r}=\pm\Big{[}\bar{h}_{rr}-\Big{(}\frac{{\rm d}\bar{R}}{{\rm d}r}\Big{)}^2\Big{]}^{1/2},
\end{eqnarray}
which provides the turning points of the embedding diagram determined by the condition ${\rm d}\bar{R}/{\rm d}r=0$, i.e. by the condition $\bar{\mathcal{Z}}_r=0$ (\ref{vel-centr-2}). Thus, also here, in the case of a fluid rotation, equatorial flow lines along which the radial component of cylindrical intensity vanishes show up as turning circles of throats or bellies of the embedding diagram. Let's highlight that concerning geodesics of the $\bar{\Sigma}_t$ hypersurface, its equatorial plane embedding diagram, and the related centrifugal force $\bar{\mathcal{Z}}^{\perp}_{\alpha}$, we get formally the same results as in the case of the $\tilde{\Sigma}_t$ hypersurface.

Along with the mentioned interpretation of the turning points, we can also reveal their direct relation to the specific enthalpy profile. Since ${\rm d} \bar{R} = {\rm d} h g_{\phi\phi}$, we find $\partial_r \ln{h} = \partial_r h/h=-\partial_r \ln{g_{\phi\phi}^{1/2}}=\partial_r w$ at the radii of the turning points. 
\section{Charged fluid around black hole in uniform magnetic field}
\label{UMF}
In this section, we construct a positively charged fluid structure rotating in the $+\phi$ direction in the Schwarzschild spacetime accompanied by an asymptotically uniform test magnetic field aligned with the direction $-z$. Such a configuration has been already studied and reviewed in our previous works \cite{Kov-etal:PRD-2014,Tro-etal:PRD-2020,Stu-etal:Uni-2020}, also with the presence of a background electric field. Here, however, the configuration plays another important role. It illustrates the basically new introduced force representations of the model in a particular case, and to preview correlations between the defined forces and their relations to embedding diagrams, which have not been presented before. The considered background fields are given by the Schwarzschild metric line element and by the only non-zero component of the vector potential  
\begin{eqnarray}
\label{metricW}
{\rm d}s^2&=&-\Big{(}1-\frac{2}{r}\Big{)}{\rm d}t^2+\Big{(}1-\frac{2}{r}\Big{)}^{-1}{\rm d}r^2+r^2\sin^2{\theta}{\rm d}\phi^2+r^2{\rm d}\theta^2,\nonumber\\
\label{waldpot}
A_{\phi}&=&\textstyle{\nicefrac{1}{2}}\,B r^2\sin^2{\theta},
\end{eqnarray}
where $|B|$ stands for the dimensionless strength of the magnetic field. 

\begin{figure}[h!]
\centering
\includegraphics[width=0.95\hsize]{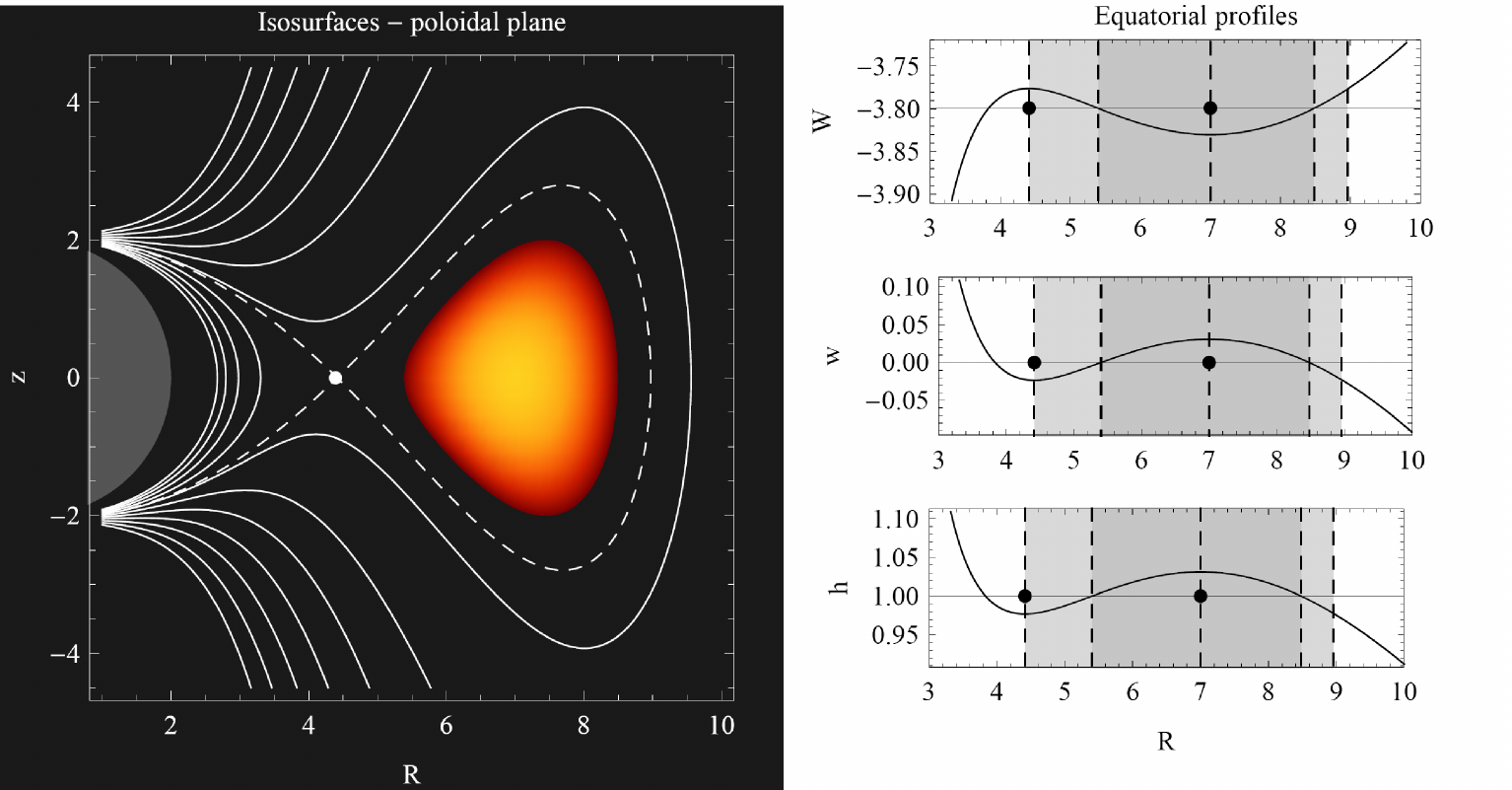}
\caption{Typical behaviour of the potential $W$, transformed pressure $w$ and specific enthalpy $h$ of the rotating charged fluid, shown in terms of their poloidal contours (poloidal sections of their isosurfaces) and equatorial profiles. The presented behaviour characterizes the rotating fluid forming toroidal structure with the profiles of $\Omega$ and $\mathcal{K}$ given by relations (\ref{Omega}). The self-crossing dashed curve corresponds to a critical isosurface with an equatorial cusp. The dots denote location of the cusp and the centre of the structure, the dashed lines denote positions of the cusp, inner edge, centre and outer edge of the structure. The shaded zones indicate the region within the critical closed isosurface (light gray), and the interior of the structure (dark gray).}
\label{Fig:1}
\end{figure}

The constructed rotating toroidal structure is shown in Fig.~\ref{Fig:1}.
It is presented in terms of the poloidal, $(R,z)$, contours and equatorial, $z=0$, profiles of $w$, related potential $W$ and specific enthalpy $h$; remind that $R=r\sin{\theta}$ and $z=r\cos\theta$. Particularly, the structure is determined by the solution of equation (\ref{united}), in terms of the Pfaffian coordinates $(\ln|U_t|,\ell,A)$ taking the form
\begin{eqnarray}
\label{samplew}
w=-\ln{|U_t|}+\frac{k}{1-k}\ln{|\ell|}+q_{0}\Big{(}\frac{2 B k}{1-k}A_{\phi}\Big{)}^{1/2}+W_{\rm ed}\equiv-W+W_{\rm ed},
\end{eqnarray}
where $\ell^2=-k g_{\phi\phi}/g_{tt}$. The integration constant $W_{\rm ed}\doteq-3.799$ corresponds to the position of edge of the structure at $r_{\rm ed}=5.4$ and  $\theta_{\rm ed}=\nicefrac{\pi}{2}$. The centre of the structure is located at $r\doteq 7.001$, corresponding to the minimum of $W$ (maximum of $w$), while the equatorial cusp is placed at $r\doteq 4.413$, corresponding to the saddle point of $W$ (maximum of the equatorial profile of $W$) and to the saddle point of $w$ (minimum of the equatorial profile of $w$); the rotational, charge and magnetic field parameters are set to $k=\nicefrac{39}{50}$, $q_0=\nicefrac{1}{6}\times 10^{17}$ and $B=-\nicefrac{6}{5}\times 10^{-17}$.        
Surface of the structure is determined by the null closed isosurface of $w$, the interior is described by the profile $w>0$. Outside this region, the solution (\ref{samplew}) is without the fluid relevance. Note that within the class of solutions (\ref{samplew}) given by the profile of the potential $W$, the structure could be extended, by setting the integration constant $W_{\rm ed}$, up to the critical isosurface of $W$ (critical equipotential surface). This critical equipotential surface     would correspond to the null isosurface of $w$ determining the most extended toroidal structure within this class of solution. Beyond the critical equipotential surface,  an equilibrium of fluid in rotation cannot be achieved.   

The solution (\ref{samplew}) corresponds to the choice of rotational regime and charge distribution characterized by the profiles     
\begin{eqnarray}
\label{Omega}
\Omega=\frac{k}{\ell},\quad \mathcal{K}=q_{0}\Big{(}\frac{\nicefrac{1}{2}B k}{1-k}A_{\phi}^{-1}\Big{)}^{1/2},
\end{eqnarray} 
in accordance with the integrability condition (\ref{icondition}). 
In more details, due to relations (\ref{OmegaUt}), (\ref{K1}), $\Omega=U^{\phi}/U^t$, $\ell=-U_{\phi}/U_t$ and $A_{\phi}=\nicefrac{1}{2}B g_{\phi\phi}$ (\ref{waldpot}), the ansatz (\ref{Omega}) determines the azimuthal four-velocity component and the specific charge distribution profiles in the form
\begin{eqnarray}
\label{q-profile}
U^{\phi}=\Big{(}\frac{\nicefrac{1}{2}B k}{1-k}A_{\phi}^{-1}\Big{)}^{1/2},\quad q=q_0\frac{\epsilon+p}{\rho}=q_0 {\rm e}^w.
\end{eqnarray}
Note that the considered profile of $q$ justifies our consideration of the equation of state in the form $\epsilon=\epsilon(p)$, required for the integrability as well;\footnote{In general, charged fluids must be characterized be the pressure, density, specific charge and energy density related as $p=p(\rho;q)$ and $\epsilon=\epsilon(\rho,p;q)$. Thus, the specific charge given by relation (\ref{q-profile}) allows for an equation of state in the form $\epsilon=\epsilon(p)$, as required for the integration.} for $p\ll\epsilon$, the specific charge distribution turns out to be almost constant, $q\approx q_0$.

The solution (\ref{samplew}) is also a solution of equation (\ref{united1}). Thus, it can be expressed equivalently in terms of the Pfaffian coordinates $(\Phi,\tilde{R},A)$ as    
\begin{eqnarray}
\label{samplew1}
w=-\Phi+\tilde{v}\ln{\tilde{R}}+q_{0}\Big{(}\frac{2 B k}{1-k}A_{\phi}\Big{)}^{1/2}+\tilde{W}_{\rm ed}\equiv-\tilde{W}+\tilde{W}_{\rm ed},
\end{eqnarray}
where $\tilde{W}_{\rm ed}\doteq-4.997$, and the rotational regime is described by the choice $v^2=\Omega\ell=k$, i.e. $\tilde{v}^2= k/(1-k)$. 

For the specific enthalpy $h$ related to our solution (\ref{samplew}), we find $\nabla_{U}(h U_{\alpha}\xi^{\alpha})=\nabla_{U}(h U_{\phi})=U^{\alpha}\partial_{\alpha}(h U_{\phi})=0$ in accordance with \cite{Rez-Zan:2013}, since $U^{\alpha}=(U^t,0,0,U^{\phi})$, $h=h(r,\theta)$ and $U_{\phi}=U_{\phi}(r,\theta)$ in our case; this is because the constructed fluid structure shares the axial symmetry with the background fields. Thus, $hU_{\phi}={\rm const}$ along fluid flow lines, whereas, in our case, different fluid flow lines are related to different constants, as it follows from the non-vanishing directional derivative  $\nabla_{V}(hU_{\phi})=V^{\alpha}\partial_{\alpha}(hU_{\phi})\neq 0$, where $V^{\alpha}=(0,V^r,V^{\theta},0)$ is a perpendicular four-vector field to the four-vector field $U^{\alpha}$. 

\begin{figure}[h!!!]
\centering
\includegraphics[width=0.5\hsize]{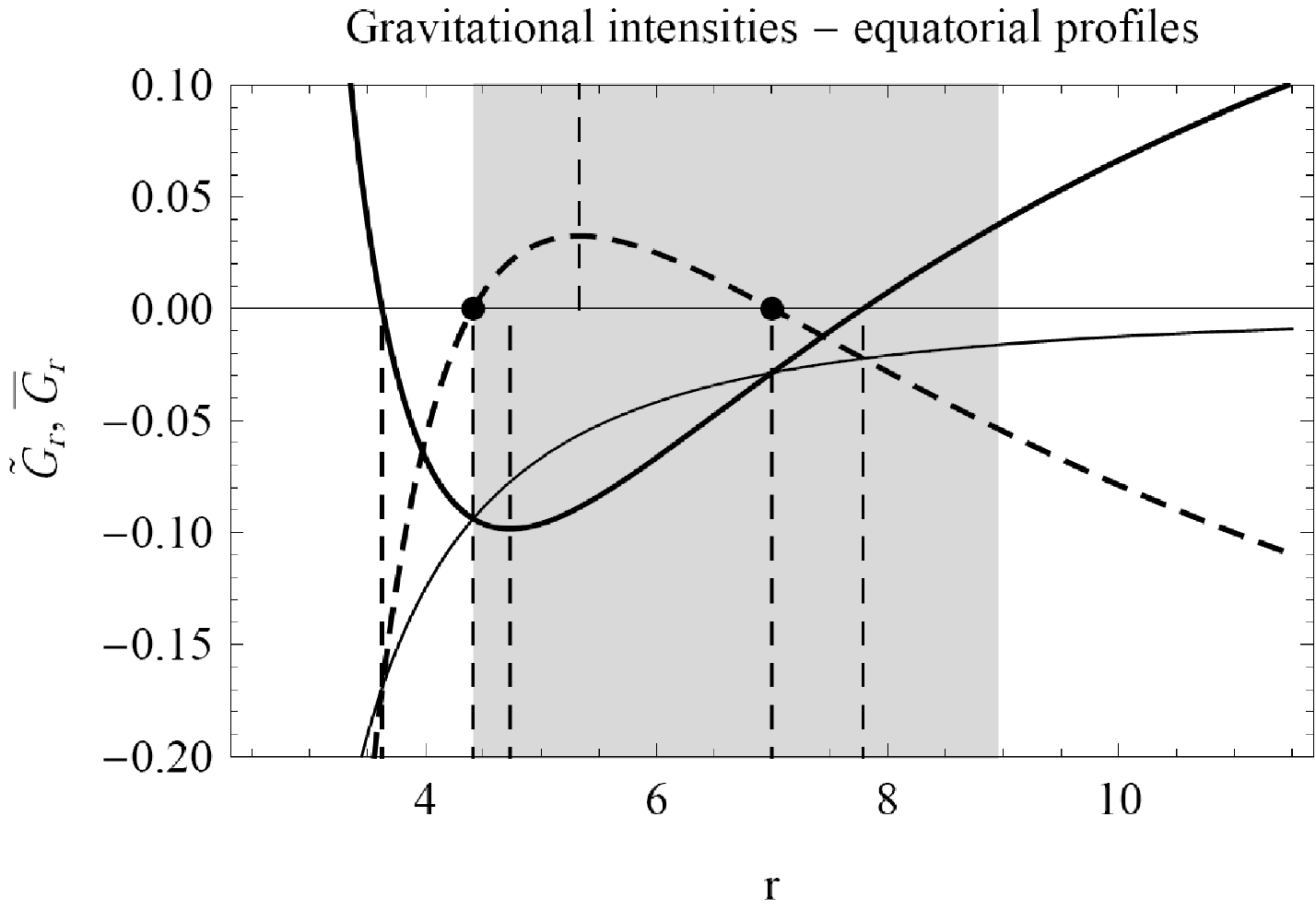}\includegraphics[width=0.5\hsize]{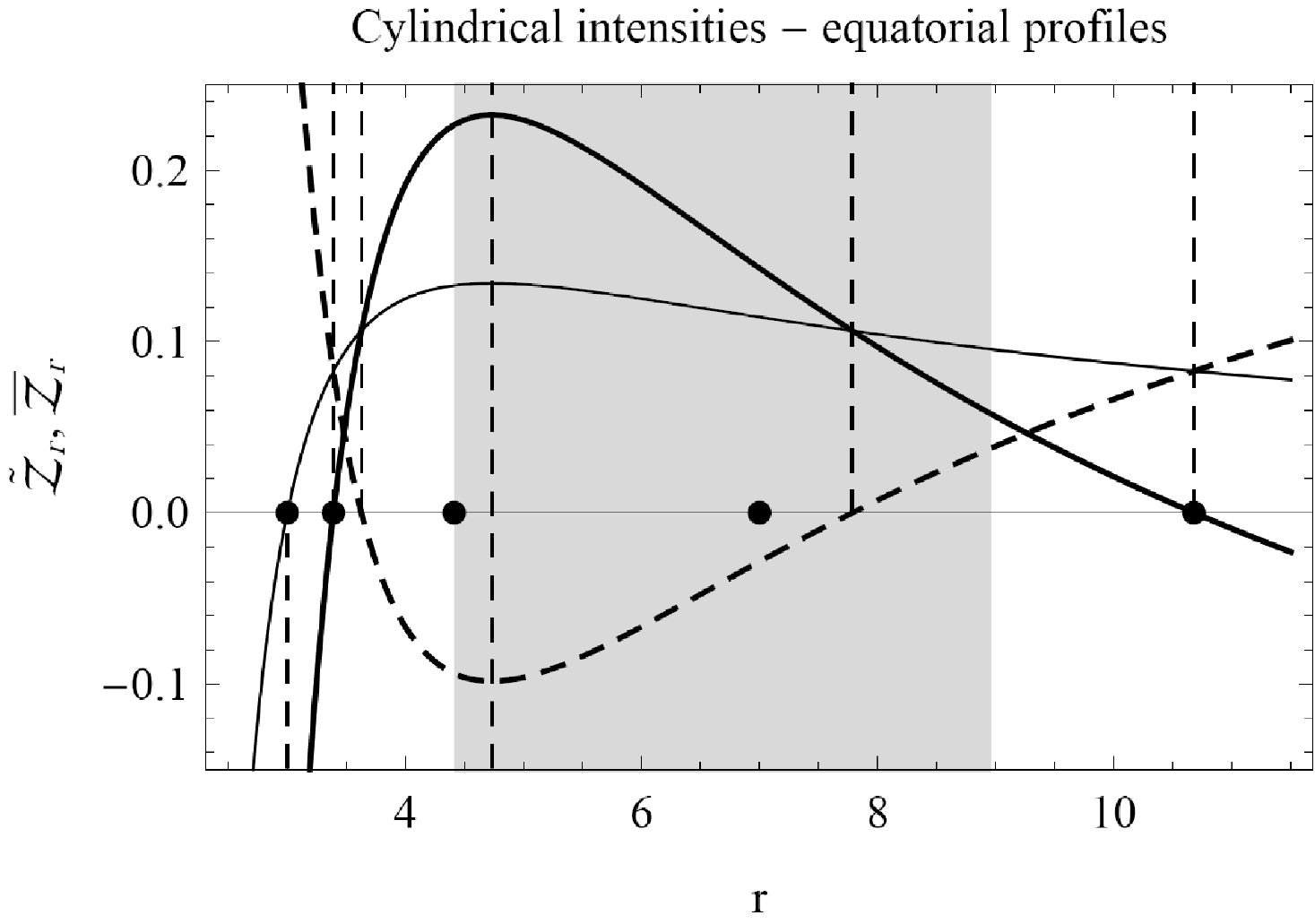}
\caption{Equatorial profiles of the radial components of gravitational forces -- gravitational intensities $\tilde{G}_r$ (left: thin solid) and $\bar{G}_r$ (left: thick solid), their difference, $\tilde{G}_r-\bar{G}_r=\partial_r w$ (left: thick dot-dashed), velocity independent parts of centrifugal forces -- cylindrical intensities $\tilde{\mathcal{Z}}_r$ (right: thin solid) and $\bar{\mathcal{Z}}_r$ (right: thick solid), and their difference, $\tilde{\mathcal{Z}}_r-\bar{\mathcal{Z}}_r=\bar{G}_r$ (right: thick dot-dashed); the dashed lines denote the radii of vanishing, equality and extrema. The dots mark positions of the cusp and the centre of structure (vanishing of $\partial_r w=0$), and of the turning points of embedding diagrams of $\tilde{\Sigma}_t$ and $\bar{\Sigma}_t$ equatorial planes (vanishing of $\tilde{\mathcal{Z}}_r$ and $\bar{\mathcal{Z}}_r$), as shown in Fig.~\ref{Fig:3}. The shaded zones indicate the region within the critical closed isosurfaces (see Fig.~\ref{Fig:1}).}
\label{Fig:2}
\end{figure}

Within an investigation of equatorial toroidal structures, it is illustrative to focus on distinctive equatorial profiles of forces, potentials, intensities and other important characteristics. Here, we survey profiles of the radial components of gravitational forces -- gravitational intensities, $\tilde{G}_r=-\partial_r \Phi$ and $\bar{G}_r=-\partial_r \Psi$, and of radial components of the velocity independent parts of centrifugal forces -- cylindrical  intensities,  $\tilde{\mathcal{Z}}_r=\partial_{r}\ln{\tilde{R}}$ and $\bar{\mathcal{Z}}_r=\partial_{r}\ln{\bar{R}}$ (see relations  (\ref{G1}), (\ref{G2}), (\ref{Z1}) and (\ref{Z2}),  and Fig.~\ref{Fig:2}); the other components are zero, as well as all the components of Coriolis forces and electric forces. At this moment, we do not map any equatorial behavior of the non-vanishing radial components of magnetic forces $\tilde{M}_r$ and $\bar{M}_r$.   
We can see that $\tilde{G}_r$ is always negative, i.e. the gravitational force can be only attractive, while $\tilde{\mathcal{Z}}_r$ changes its sign at the radius $r=3$ (radius of its vanishing and radius of the photon sphere of the Schwarzschild spacetime), i.e. the centrifugal force $\tilde{Z}_r$ can be repulsive above this radius and attractive below, i.e. close to the event horizon; this non-Newtonian attractive centrifugal action has been addressed in literature many times \cite{Abr-Pra:MNRAS-1990,Son-Mas:MNRAS-1996}. On the other hand, $\bar{G}_r$ changes its signs  at two radii, $r\doteq 3.627$ and $r\doteq 7.788$, i.e. the newly defined gravitational force can be repulsive close to the event horizon. Here, the centrifugal force $\bar{Z}_r$ can be attractive, but changing its orientation twice, at the radii $r\doteq 3.391$ and $r\doteq 10.679$ (radii of its vanishing), it can be attractive also further from the horizon, in contrast to the force $\tilde{Z}_r$. 
Moreover, since there is $\tilde{G}_r-\bar{G}_r=\partial_r w$ (\ref{Psi-pot}), $\tilde{G}_r=\bar{G}_r$ at the radii corresponding to positions of the cusp and the centre of structure. Moreover, we can see that $\tilde{\mathcal{Z}}_r-\bar{\mathcal{Z}}_r=\bar{G}_r$.

\begin{figure}[h!]
\centering
\includegraphics[width=0.35\hsize]{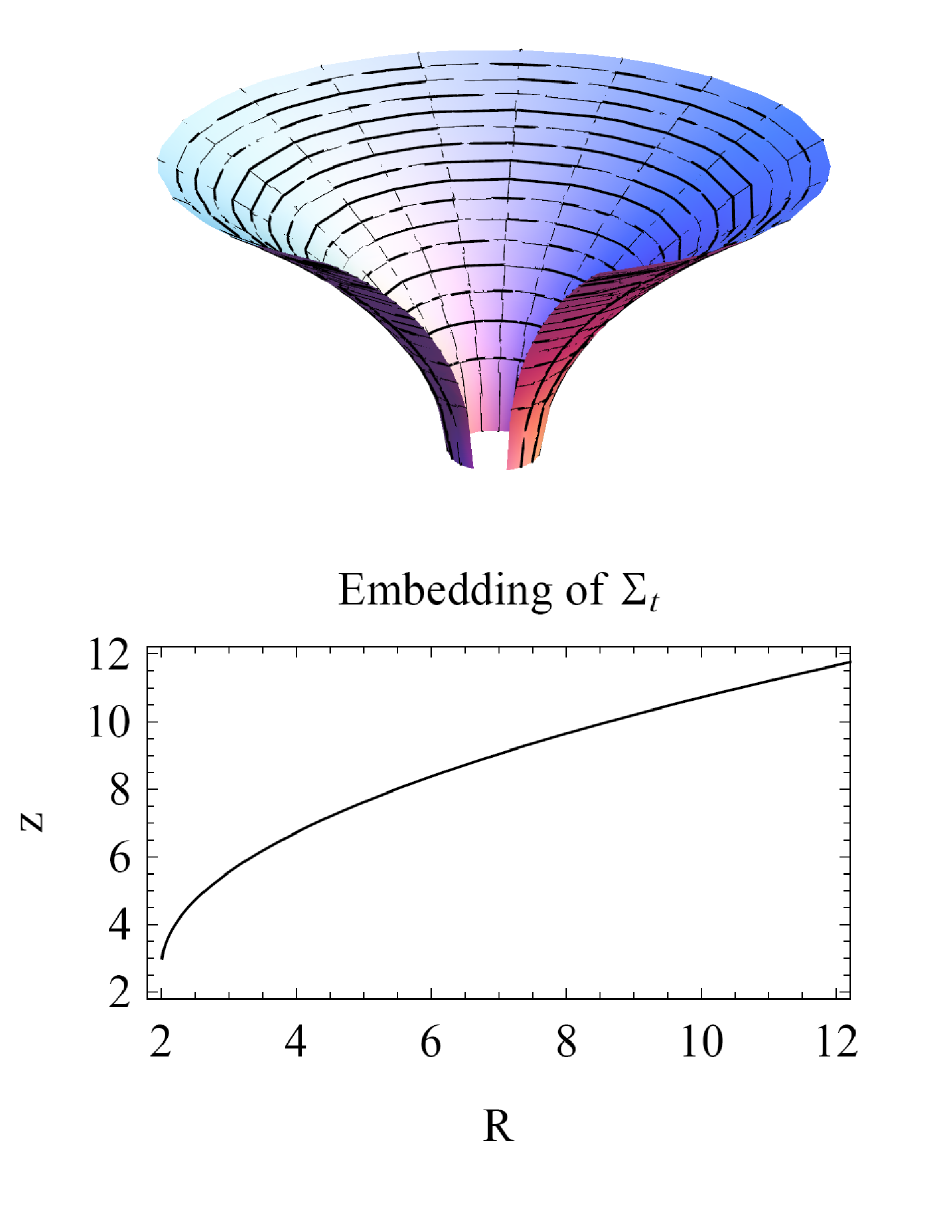}\includegraphics[width=0.35\hsize]{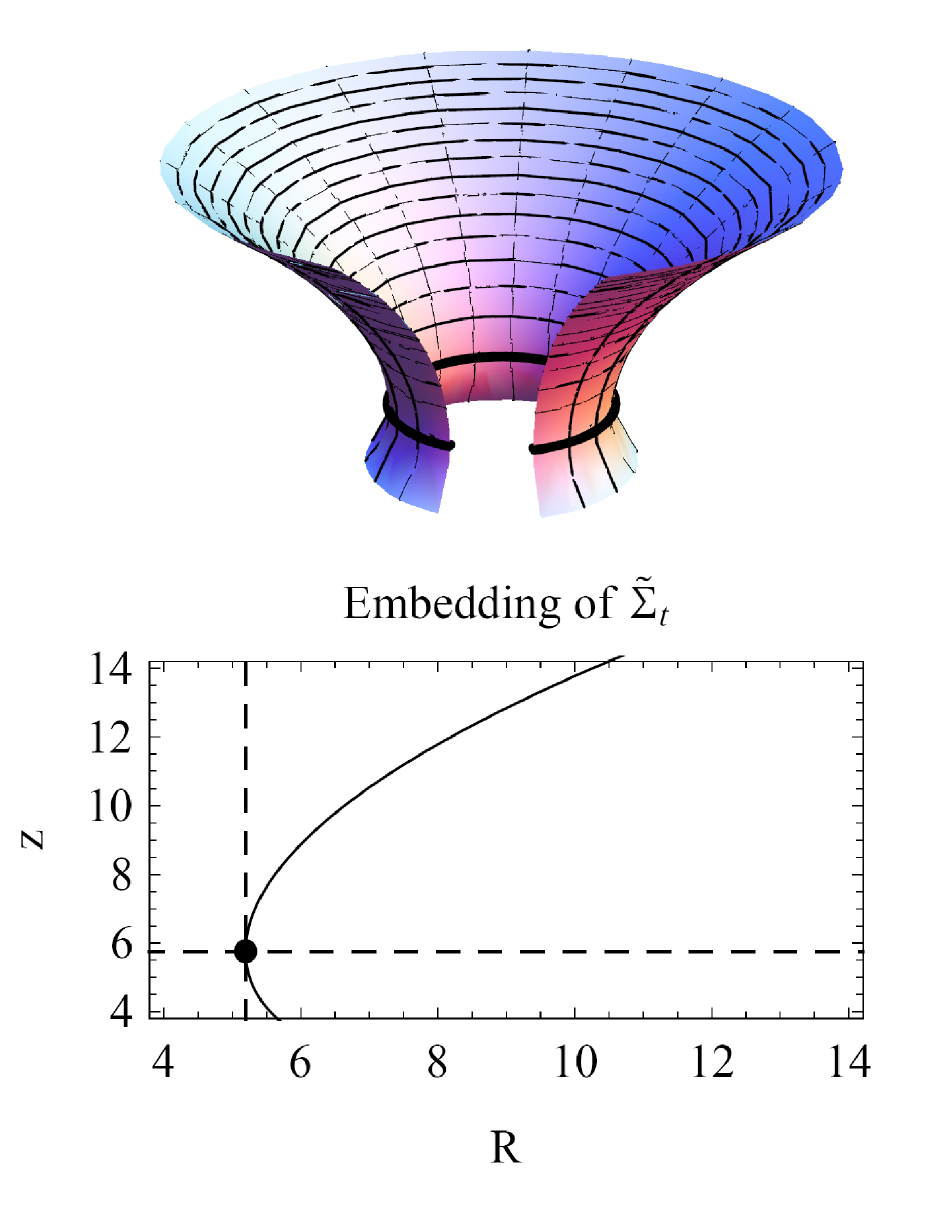}\includegraphics[width=0.35\hsize]{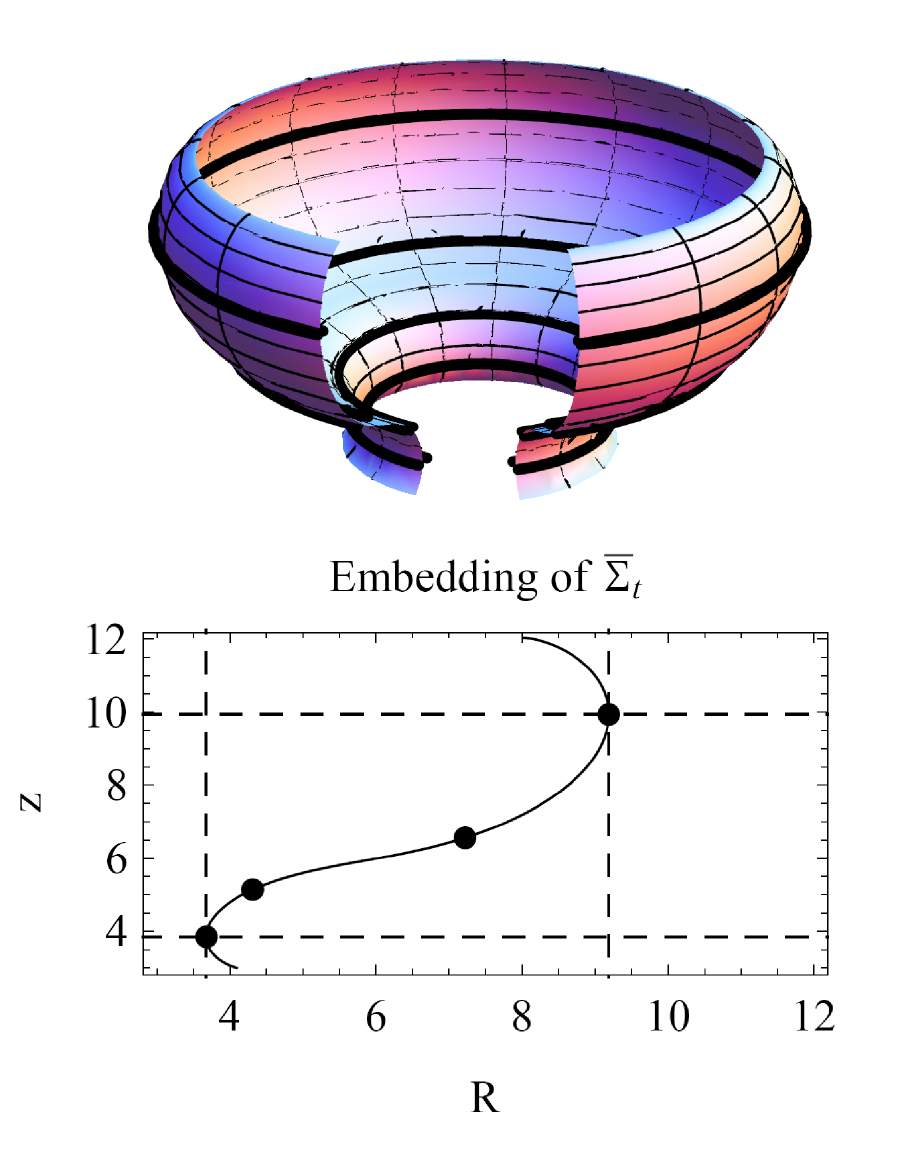}
\caption{Embedding diagrams of $\Sigma_t$, $\tilde{\Sigma}_t$ and $\bar{\Sigma}_t$ equatorial planes and their poloidal profiles. Turning points of the diagrams correspond to the radii where $\tilde{\mathcal{Z}}_r=0$ and $\bar{\mathcal{Z}}_r=0$ (see Fig.~\ref{Fig:2}), as denoted by the dots and dashed lines; the dots mark also positions of the cusp and the centre of structure. The embedding diagram of the $\Sigma_t$ equatorial plane is monotonous.}
\label{Fig:3}
\end{figure}

Behaviour of the cylindrical intensities $\tilde{\mathcal{Z}}_r$ and $\bar{\mathcal{Z}}_r$ is directly reflected in shapes of the embedding diagrams of $\tilde{\Sigma}_t$ and $\bar{\Sigma}_t$ equatorial planes, assigning $R^2=\tilde{h}_{\phi\phi}=\tilde{R}^2$ and $R^2=\bar{h}_{\phi\phi}=\bar{R}^2$, respectively; turning points of the embedding diagrams correspond to radii of the cylindrical intensities vanishing (see Fig.~\ref{Fig:3}).  
In the case of the $\bar{\Sigma}_t$ embedding diagram, we find for the specific enthalpy the relation $\partial_r \ln{h} = \partial_r h/h=-1/r=\partial_r w$ at the radii of the turning points; here, the gradient of the specific enthalpy logarithm is directly determined only by the negative reciprocal of the turning points radii.

\subsection{Comment on astrophysical contextualization}
Physically well-defined background fields (\ref{waldpot}) were chosen for their unique simplicity, which is very convenient for the analytical treatment and clear illustration of the force based model. Besides, such a composition of fields is usually applied as an astrophysically relevant simplification. 

At first, note that the chosen electromagnetic field corresponds to the non-rotating and uncharged limit of the `Wald solution' \cite{Wal:PRD-1974} of Maxwell equations (assumed with the Kerr metric) -- the test-field solution describing an electromagnetic field around `weakly' charged rotating black hole embedded in a `weak' asymptotically uniform magnetic field; our limit is then suitable for a static uncharged black hole in the uniform magnetic field. 

Consequently, the Wald solution turns out to be reliable for description of dynamics of matter around stellar-mass black holes (${\rm M}_{\odot}\lesssim M \lesssim 10^2 {\rm M}_{\odot}$) embedded in the uniform galactic magnetic field. This field is sufficiently uniform and strong enough, $|B_{\rm SI}| \lesssim 10^{-7}{\rm T}$ \cite{Cro-etal:Nat-2010}, to influence significantly the motion of charged particles characterized not only by the high specific charges, $q\sim 10^{18}$, characteristic for cations, but also to those much lower specific charges, characteristic for charged dust grains (see, e.g. \cite{Kov-Kop-Kar-Stu:CQG-2010,Kop-Kar-Kov-Stu:APJ-2010,Shio-etal:PRD-2014,Tur-etal:PRD-2016,Kol-Tur-Stu:EPJC-2017}). Adopting this astrophysical scenario also for our illustrative example, for $M\sim 10^2{\rm M}_{\odot}$, the chosen $B=-\nicefrac{6}{5}\times 10^{-17}$ corresponds to $B_{\rm SI}\sim - 10^{-4} {\rm T}$, which is beyond the galactic field limit $|B_{\rm SI}| \sim 10^{-7}{\rm T}$. However, for an intermediate black hole with a mass $M\sim 10^5{\rm M}_{\odot}$, we get the satisfactory value $B_{\rm SI}\sim - 10^{-7}{\rm T}$; the chosen $q_0=\nicefrac{1}{6}\times 10^{17}$ then corresponds to the specific charge in centre of the structure $q\sim 10^{16}$independent on the considered mass of the central black hole. 

It is important to note that due to the considered test character of the magnetic field the parameter $B$ does not influence the metric. Together with the parameter $q_0$ it only scales the electromagnetic term, both being coupled in the only electromagnetic parameter, $\mathcal{B}=q_0 B$. For the purpose of direct physical interpretation, however, we keep the parameters $q_0$ and $B$ separated in all relations, even if only their product $\mathcal{B}$ influences the fluid structures. Consequently, the structure presented above corresponds to all the combinations of the external magnetic field strength and magnitude of the specific charge determined by the chosen value $\mathcal{B}=-\nicefrac{1}{5}$; the mutual change of the signs of $B$ and $q_0$, not affecting the value of $\mathcal{B}$, allows us to interpret the constructed structure also as the one consisting of negatively charged fluid rotating in the $+z$ sense oriented asymptotically uniform magnetic field. Moreover, by adopting the 
Wald solution, we are not limited only to the consideration of the very weak galactic magnetic field. Another relevant astrophysical scenarios with much stronger, but still test, uniform magnetic fields can be considered \cite{Kov-etal:PRD-2014,Stu-Kol:EPJC-2016,Tur-etal:APJ-2020}.       

In general, an astrophysical contextualization and discussion of the charge fluid model is a very complex problem, dependent on the chosen background fields, their parameters, but also on the fluid parameters. Moreover, one must take into account the restrictive conditions of the model (low mass and weakly charged structures, etc.). The detailed discussion of the astrophysical relevance of the charge fluid structures constructed in the considered generalized background can be found in \cite{Kov-etal:PRD-2014,Tro-etal:PRD-2018}.
\section{Conclusions}
We have shown that a rotation of charged perfect fluid in given gravitational and electromagnetic fields of a compact object can be described within two covariant force representations. The first representation is based on a definition of hypersurface forces, following from a proper four-acceleration hypersurface projection and its specific decomposition; the second one is additionally enhanced by a specific enthalpy profile of the rotating fluid. 

In contrast to the description of rotating charged fluid in terms of the momentum conservation equation -- the Euler equation with a charge term, written in a standard representation, we found here the rotating fluid described by inertial forces, such as the velocity and charge independent gravitational forces, the velocity dependent and charge independent centrifugal and Coriolis forces, and by real forces, such as the charge and velocity dependent electric and magnetic forces, conveying the electromagnetic interaction. Apart from a debatable Newtonian-like interpretation of the used force formalism, the applied decompositions are useful for astrophysical studies in axially symmetric and stationary spacetimes, where they provide well arranged meaningful terms. Especially, it has turned out that searching for a particular class of analytical solutions of the Euler equation can be more convenient and straightforward if the decompositions are employed.

It is also nice to see how the force representations are reflected in geometries of related conformal spacetimes and their hypersurfaces. This can be effectively visualized by embedding diagrams of equatorial planes of these hypersurfaces, being well characterized by their turning points mapping the radii of equatorial rotating fluid flow lines along which the radial components of velocity independent parts of centrifugal forces -- cylindrical intensities, vanish. 
In general, fluid flow lines in a special ordinary hypersurface along which the velocity independent parts of centrifugal forces vanish are mapped by 
geodesics of both the related conformal hypersurfaces. 
Moreover, neutral fluid flow world-lines from an ordinary spacetime are mapped by geodesics of the related conformal spacetime with its geometry metric rescaled by a factor containing a specific enthalpy profile of the fluid, possibly forming a rotating toroidal structure. Charged fluid flow world-lines are mapped by world-lines determined by the charged test particles equation of motion -- the Lorentz-like equation, in the related conformal spacetime.

In general, the introduced force representations of the model have provided us with theoretical frameworks for fluid flow investigations that are alternative to the standard one applied in the previous works \cite{Kov-etal:PRD-2011,Kov-etal:PRD-2014,Kov-etal:PRD-2016,Tro-etal:PRD-2018,Schr-etal:PRD-2018,Stu-etal:Uni-2020,Tro-etal:PRD-2020}. Due to the related conformal geometries, we have also obtained interesting mathematical contextualization and perspectives on the emerging toroidal structures. These can be very useful for further astrophysically motivated studies, since the toroidal flows of fluids could serve as good approximation to actual flows during accretion processes, exactly as periodic orbits do to represent segments of actual orbits \cite{Lev-Per:PRD-2008}.

\ack
The authors JK, PS and ZS would like to express their acknowledgement for the institutional support of the Research Centre for Theoretical Physics and Astrophysics, Institute of Physics, Silesian University in Opava. JK and VK thank for the support from the project of the Czech Science Foundation GA\v{C}R ref. 19-01137J. ZS thanks for the support from the project of the Czech Science Foundation GA\v{C}R ref. 19-03950S. YK thanks for the support from JSPS KAKENHI Grant Number JP19K03850. Special thanks go to dr. Petr Blaschke (Silesian University in Opava) for a fruitful discussion.

\section*{References}

\bibliographystyle{unsrt}

\bibliography{Kovar-bibl-1}

\end{document}